# Equilibrium protein adsorption on nanometric vegetable-oil hybrid film/water interface using neutron reflectometry

Antigoni Theodoratou,[1,4*] Lay-Theng Lee,[2] Julian Oberdisse,[3*] Anne Aubert-Pouëssel[1]

[1] Institut Charles Gerhardt Montpellier (ICGM), UMR5253 CNRS-UM-ENSCM, Place Eugène Bataillon, 34090 Montpellier, France.

[2] Laboratoire Léon Brillouin, CEA-CNRS, CEA Saclay, Université Paris-Saclay, 91191 Gif-sur-Yvette, France.

[3] Laboratoire Charles Coulomb (L2C), UMR5221 CNRS-UM, Place Eugène Bataillon, 34090 Montpellier, France.

[4] European Institute of Membranes (IEM), UMR5635 CNRS-ESNCM, 300 Avenue du Professeur Emile Jeanbrau, 34090 Montpellier, France.

*E-mail: antigoni.theodoratou@umontpellier.fr, julian.oberdisse@umontpellier.fr

## ABSTRACT:

Nanofilms of thickness of about two nanometers have been formed at the air-water interface using functionalized castor oil (ICO) with cross-linkable silylated groups. These hybrid films represent excellent candidates for replacing conventional polymeric materials in biomedical applications, but they need to be optimized in terms of biocompatibility which is highly related to protein adsorption. Neutron reflectivity has been used to study the adsorption of two model proteins, bovine serum albumin and lysozyme, at the silylated oil (ICO)-water interface in the absence and presence of salt at physiologic ionic strength and pH and at different protein concentrations. These measurements are compared to adsorption at the air-water interface. While salt enhances adsorption by a similar degree at the air-water and the oil-water interface, the impact of the oil film is significant, with adsorption at the oil-water interface three- to four-fold higher compared to the air-water interface. Under these conditions, the concentration profiles of the adsorbed layers for both proteins indicate multilayer adsorption: The thickness of the outer layer (oil-side) is close to the dimension of the minor axis of the protein molecule, ~ 30 Å, suggesting a side-way orientation with the long axis parallel to the interface. The inner layer extends to 55 - 60 Å. Interestingly, in all cases, the composition of oil film remains intact without significant protein penetration into the film. The optimal adsorption on these nanofilms, 1.7 - 2.0 mg·m$^{-2}$, is comparable to the results obtained recently on thick solid cross-linked films using quartz crystal microbalance and atomic force microscopy, showing in particular that adsorption at these ICO film interfaces under standard physiological conditions is non-specific. These results furnish useful information towards the elaboration of vegetable oil-based nanofilms, in direct nanoscale applications or as precursor films in the fabrication of thicker macroscopic films for biomedical applications.

## INTRODUCTION

The design of new biomaterials is driven by applications, in particular medical applications in orthopedics (joints, ligaments),[1] dentistry,[2] pharmacy (drug delivery systems and wound healing)[3] and tissue engineering.[4] The synthesis of polymeric materials from renewable sources like castor oil is a promising route towards bio-sourced coatings for biomedical applications, because of their low levels of toxicity, biodegradability and inexpensiveness. Moreover, castor oil-based materials have shown enhanced cytocompatibility, biocompatibility in living tissue and biodegradability.[5-6]

Castor oil is produced from the *Ricinus communis* plant, which belongs to the *Euphorbiaceae* family. It is non-edible, and its main component is ricinoleic acid, which contains hydroxyl groups favoring both solubilization of lipophilic drugs, and chemical functionalization. Due to the mechanical requirements of any coating, a cross-linked polymer network has to be formed, if possible, of tunable and enhanced hardness. We have shown recently that this can be achieved by introducing a silica precursor, which reacts with the hydroxyl groups of the oil molecule and generates cross-links. This reaction leads to the formation of a solid, eco-friendly polymeric material.[7]

The adsorption properties of biopolymer coatings are crucial for medical applications. For this reason, the understanding of interactions of biological materials at a nanoscale has been a topic of several studies.[8-9] We have recently studied the structure, mechanical properties and cytotoxicity on *thick* (hundreds of microns) films[7]. Previously, we explored the formation of *intermediate thickness* (hundreds of nanometers) films obtained by spin-coating and cross-linking functionalized castor oil (ICO), and investigated their tunability and protein adsorption from phosphate buffer saline (PBS) of the model proteins, BSA and lysozyme.[10] In that study, atomic force microscopy (AFM) and quartz crystal microbalance with dissipation (QCM-D) were used to follow adsorption/desorption kinetics. The surface coverage in aqueous medium was found for both proteins to be close to $30 \pm 8$ %, hence showing that the major part of the substrates is occupied by the phosphate buffer. For this coverage, the thickness of both proteins was evaluated with QCM-D and AFM. It was found to be, for BSA, $d_{AFM} = 6.5 \pm 5.4$ nm and $d_{QCM-D} = 8.3 \pm 1.1$ nm, and for lysozyme $d_{AFM} = 7.9 \pm 5.5$ nm and $d_{QCM-D} = 7.6 \pm 0.5$ nm. These results will be compared below with the coverage and thickness obtained for lysozyme and BSA at equilibrium adsorption on *very thin nanofilms* using neutron reflectometry.

When proteins adsorb at the air-water interface they adopt a thinner conformation comparing to their 'native' conformation in the bulk. According to Wierenga et al.[11] and Yano et al.,[12] during adsorption unfolding takes place and this is due to the similar or slower kinetics of adsorption comparing to the kinetics of unfolding. During protein adsorption at the air-water interface the proteins lose their biological activity leading to protein denaturation. The conformational changes of proteins during the adsorption are accompanied by binding of the hydrophobic groups on the interface.[13] In general, the parameters that can control the adsorption are (a) the pH, (b) the concentration, (c) the presence of salts in the solution (d) and the temperature.[14] The adsorption rate can be also ruled by the flexibility of the polypeptide chain and the hydrophobicity/hydrophilicity of the protein surface.[15] One may note that protein denaturation at fluid interfaces can be studied in presence of chemical denaturants.[16] X-ray and neutron reflectometry have shown that the presence of chemical denaturant (guanidinium hydrochloride) during lysozyme adsorption at the air-water interface affects the orientation of the protein with no significant distortion of the tertiary structure of the protein, at low denaturant concentrations (up to 2 mol·L$^{-1}$).

It has been shown that the presence of oil plays an important role for the proteins to adsorb at the oil-water interface. Protein adsorption increases when the hydrophobicity of the oil layer is increased. Recently, the effect of oil on protein adsorption was investigated by Bergfreund et al. showing that the polarity of the oil molecule as well as its hydrophobicity and chain length can control the final properties of the protein interfacial layers.[17] The authors showed that β- lactoglobulin forms networks with an identical viscoelastic response, as measured with interfacial shear rheology, at several n-alkane interfaces. Moreover, comparing the viscoelastic properties of β-lactoglobulin films at the hexane/water and air/water interface, the authors found that at the hexane/water interface stronger films can be formed. Campana et al. showed that BSA at the hexadecane-water interface has a compact arrangement of molecules[18]. Moreover, the presence of oil may cause denaturation of the protein. A description of the changes of the secondary structures of BSA when they are adsorbed at the oil-water interfaces has been reported recently by Day et al. [19] It was shown by synchrotron radiation circular dichroism that both BSA and lysozyme when adsorbed at oil-water interfaces undergo a systematic reduction in the α-helical content and an increase in the β-sheet content. Also, BSA undergoes much larger conformational re-arrangements at interfaces comparing to lysozyme, due to its higher number of disulfide bonds.[19] Mitropoulos et al. have related the adsorption of globular proteins to their thermodynamic stability and free energy of unfolding; for example proteins like lysozyme with high thermodynamic stability can adsorb at the interface with a low rate but can eventually form very stable networks.[20] Most of the literature of protein adsorption is focused on the changes of their secondary structure, while the spatial orientation of proteins remains a topic of investigation, either at flat interfaces as studied here, or on curved surfaces of nanoparticles. [21-24]

*Very thin* castor oil films of typically nanometric thickness may be of importance in medical applications.[25] In this paper, a neutron reflectometry (NR) study of protein adsorption (BSA and lysozyme) on castor oil films of nanometric thickness, in the absence and presence of salt at physiological ionic strength, under equilibrium conditions in the absence of flow is presented. The resulting adsorption will also be compared to previously obtained results from QCM-D measurements and from AFM which gives access in direct space to surface coverage and thickness, however on much smaller pieces of sample.

## MATERIALS AND METHODS

**Materials.** Pharmaceutical grade castor oil (CO; average MW = 934 g·mol$^{-1}$) was purchased from Cooper Pharmaceutique. 3-(Triethoxysilyl) propyl isocyanate (IPTES; MW = 247.4 g·mol$^{-1}$), dibutyltin dilaurate (DBTDL; MW = 631.6 g·mol$^{-1}$) and Deuterium oxide (D$_2$O) were supplied by Sigma-Aldrich and used as received. Bovine serum albumin (BSA, MW 66 kDa, isoelectric point (pI) 4.7) and lysozyme (LYZ, MW 14.3 kDa, pI 11.4) from chicken egg white were bought from Sigma Aldrich.

**Synthesis of functionalized castor oil (ICO).** A solvent-free reaction was used to functionalize castor oil with a silica precursor (IPTES) as coupling agent to obtain castor oil with cross-linkable functions. The reaction was performed at 60 °C for 30 min under magnetic stirring in the presence of a catalyst (0.8 % w/w of DBTDL). The functionalization degree, molar ratio of castor oil to silica precursor to hydroxyl groups (OH), was equal to 1.

**Film preparation at the air-water interface.** When ICO is spread at the air-water interface, a sol-gel reaction occurs via a hydrolysis and condensation step leading to the formation of a hybrid film. The kinetics of the cross-linking reaction of thick films (200 μm), as well as the physico-chemical characterization of the films have been reported recently.[7] In the current study, we have used the same approach to create nanometric films at the air-water interface. The ICO oil was diluted in chloroform to a concentration of c = 5 mg·mL$^{-1}$. For each experiment, 40 μL of this ICO solution was spread using a micro-syringe on the protein solution surface. Two proteins dissolved in $D_2O$ at two ionic strengths and two bulk concentrations, $C_p$ were studied at the air-$D_2O$ and (ICO) oil-$D_2O$ interface (see Supporting Information, SI, Figure S1): BSA and LYZ in pure $D_2O$ and in 155 mM NaCl in $D_2O$, in the absence and presence of spread oil, and at protein bulk concentration $C_p$ = 1 mg·mL$^{-1}$ and 5 mg·mL$^{-1}$. All the solutions were freshly-prepared, stored in the refrigerator and used within 2 days. The pD of the protein solutions were 7.0-7.2, determined by pH measurements with a Mettler Toledo pHmeter Seven compact equipped with a microelectrode. The values were associated to the standard correction pD=pH+0.4.[26]

**Methods.** Specular neutron reflectivity experiments were performed on the time-of-flight neutron reflectometer Hermes at the ORPHEE reactor (Laboratoire Léon Brillouin, CEA-Saclay) using a polychromatic beam with wavelength λ ≈ 2.5 to 25 Å. The horizontal beam was bent using a supermirror onto the liquid surface at a grazing incident angle $\theta$ = 1.6° (angular resolution $\delta\theta/\theta$ = 0.07) giving a corresponding momentum transfer range $q = 4\pi\sin\theta/\lambda$ from about 0.014 to 0.14 Å$^{-1}$. The liquid sample container was a Teflon cell with a surface area of 9.3 x 3.6 cm that was positioned on an aluminum cell attached to an anti-vibration support. After introducing 12 mL of the protein solution (and after spreading the oil), the sample cell was closed hermetically with an aluminum cover with quartz windows to allow the neutron beam to pass through with minimal absorption. This set-up allowed measurements of liquid surfaces over an extended period of time without loss of liquid by evaporation. All measurements were carried out at room temperature, T = 22 - 23 °C. Reflectivity spectra were acquired at two-hour intervals. For these samples, no changes in kinetics were observed after 4h and the subsequent spectra were summed for improved statistics.

**Fitting procedure and data analysis.** The reflectivity curves were fitted with a model of *n*-layers of constant scattering length density, $\rho$. In the absence of absorption, the scattering length density of a material is defined as $\rho = N_A \sum (\delta_j/A_j) b_j$ where $N_A$ is the Avogadro's number, and $\delta_j$, $A_j$ and $b_j$ are the mass density, atomic weight and coherent scattering length of atom $j$ respectively. For background subtraction, we have used the constant value of $2 \times 10^{-6}$, which corresponds to the incoherent signal of deuterium, determined at high Q. For the $n$-layer model, the scattering length density profile normal to the surface is given by:

$$\rho(z) = \sum_{n=0}^{N} \left( \frac{\rho_i - \rho_{i+1}}{2} \right) \left( 1 - erf \frac{(z-z_i)}{\sigma_i} \right) \tag{1}$$

The adjustable parameters are the scattering length density $\rho_i$ and the thickness $d_i = z_i - z_{i+1}$ of layer $i$, and $\sigma_i$, the interfacial roughness between layer $i$ and $i+1$ described by an error function. $n = 0$ and $n = N + 1$ represent the two semi-infinite media air ($\rho_0 = 0$) and the $D_2O$ subphase ($\rho_{N+1} = 6.2 \times 10^{-6} \text{ Å}^{-2}$) respectively. The best-fit parameters were determined by comparing the experimental and calculated curves by iteration and minimization of the $\chi^2$ parameter. From the best-fit scattering length density profile, $\rho(z)$, the protein volume fraction $\phi_p$ can be extracted from the relation:

$$\rho = \phi_p \rho_p + (1 - \phi_p) \rho_s \tag{2}$$

where $\rho_p$ and $\rho_s$ are the scattering length densities of the protein and the solvent respectively, and $\phi_p$ the protein volume fraction. The scattering length densities for BSA and lysozyme used are $\rho_p = 3.3 \times 10^{-6}$ and $3.66 \times 10^{-6} \text{ Å}^{-2}$ respectively [14, 27-28] and the scattering length density of the solvent is $\rho_s \approx 6.2 \times 10^{-6} \text{ Å}^{-2}$ but can vary slightly depending on the extent of exposure of the sample to air due to exchange of the $D_2O$ with $H_2O$ from the air. This small variation in the substrate scattering length density does not pose a problem because $\rho_s$ is extracted directly from the critical total reflection edge, $q_c$ of the reflectivity curve from the relation: $q_c = 4\sqrt{\pi \rho_s}$ , and the signals from the adsorbed layers are located far from $q_c$ . For ICO, its scattering length density is calculated to be $\rho_{ICO} = 0.57 \times 10^{-6} \text{ Å}^{-2}$ taking the bulk density of the oil, $\delta_{ICO} = 0.961$ g/cm$^3$.[29] It is to be noted that this $n$-layer model represents $n$ regions of constant scattering length density and does not necessarily equate to $n$ molecular layers. From the concentration profile, the total adsorbed quantity, $\Gamma$ in mg·m$^{-2}$ can be evaluated from:

$$\Gamma = \delta_p \times \phi_p \times d \qquad (3)$$

$\delta_p$ is the density of the protein, $d$ the thickness of the layer, and $\phi_p$ the protein volume fraction deduced from Eq. (2). The protein densities are taken to be $\delta_p = 1.40$ g/cm$^3$ for BSA[14] and 1.43 g/cm$^3$ for lysozyme.[28] For an *n*-layer profile, the total adsorption density is the integral of all contributing layers following Eq. (3) for each layer.

## RESULTS AND DISCUSSION

By fitting the reflectivity spectra, we have used an *n*-layer model of constant scattering length density in each layer. We start with a didactical comparison of reflectometry curves measured for a sequence of samples of increasing number of model layers and parameters. Figure 1a shows the reflectivity spectra for the solvent alone at the air-D$_2$O (155 mM NaCl) interface, adsorbed lysozyme at the air-D$_2$O (155 mM NaCl) interface, and adsorbed lysozyme in the presence of oil, at the ICO-D$_2$O (155 mM NaCl) interface. These spectra are plotted as a function of $R * q^4$ versus $q$ to emphasize the high-$q$ region. In this representation, the peak situated at about 0.02 Å is the critical total reflection edge $q_c$ mentioned above from which the scattering length density of the subphase is deduced. Qualitatively, due the lower scattering length densities of the proteins and the oil compared to the scattering length density of D$_2$O, the presence of adsorbed proteins and of oil layer decreases the reflectivity with respect to the D$_2$O subphase alone. Therefore, the higher the protein adsorption, the lower the reflectivity $R$. The solid lines are the best-fit curves with corresponding scattering length density profiles shown in Figure 1b. For lysozyme at the air-D$_2$O (155 mM NaCl) interface, the signal is low with respect to the solvent alone, and a 1-layer profile is adequate to fit the curve; increase in the number of layers does not improve the quality of the fit as indicated by the $\chi^2$ value. On the other hand, in the presence of oil, at the ICO-D$_2$O (155 mM NaCl) interface, the signal is increased significantly; here, a 3-layer model is required to give a satisfactory fit to the protein layer. The fitted parameters are the scattering length density $\rho$ and the thickness $d$ of each layer, and the roughness $\sigma$ between each layer. The protein concentration in each layer is evaluated from Eq. (2) and the total adsorption density from Eq. (3). In the presence of the spread oil layer, the ICO layer is also fitted accordingly but the protein concentration is extracted only from scattering densities of the layers below the oil layer, as indicated in the schematic in Figure 1c. This assumes non-mixing of protein and oil, but it will be seen later that in most cases, the ICO layer appears to keep its integrity, as indicated by the value of the fitted scattering length density that remains almost constant and close to that for pure ICO.

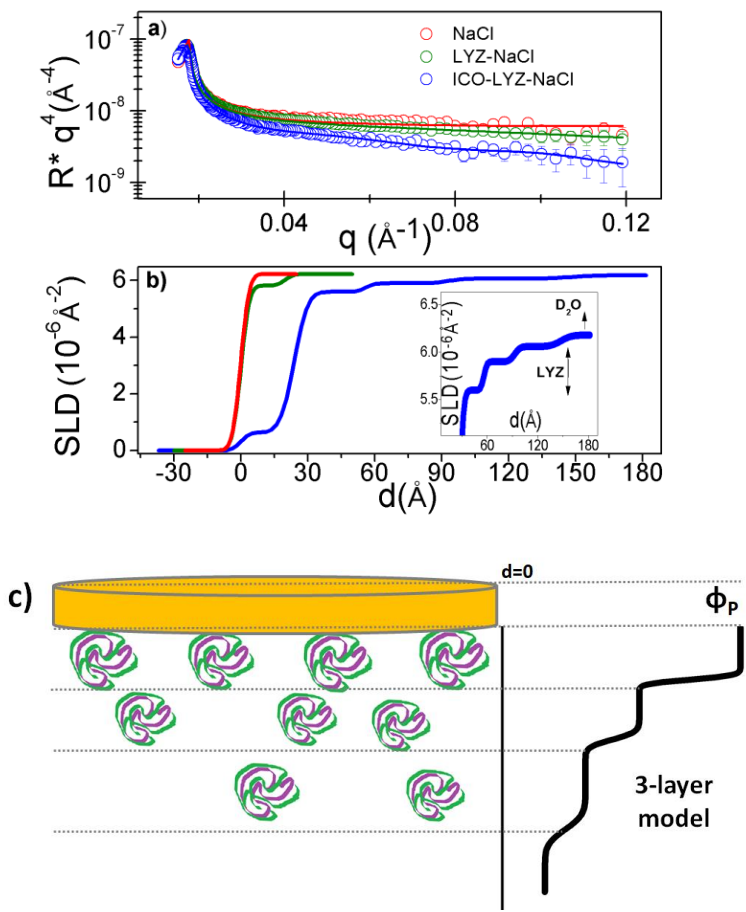

**Figure 1:** Neutron reflectivity curves plotted as as function of $R * q^4$ versus $q$ (a) for the adsorption of lysozyme at the air-$D_2O$ (green) and ICO-$D_2O$ (blue) interface in the presence of 155 mM NaCl; the curve for the solvent alone is also shown (red). The solid lines are best-fit curves with the corresponding scattering length density profiles (SLD) shown in (b); a zoom for the ICO-$D_2O$ interface is shown in the inset. In (c) a schematic representation of the protein layer model accompanied by its volume fraction profile $\phi_p(d)$ is shown.

In order to address the degree of the cross-linking in the ICO films during the neutron reflectivity experiments we have performed infrared spectroscopy measurements on nanometric films (almost 100-200 nm of thickness), that indicated that after 8 hours the state of the film is very close to the state at which the film is fully cross-linked (by means of thermal curing), and thus in a solid state. The data of this measurement can be found in the SI (Figure S3). Consequently, since the spread ICO films in our experiments are 2 - 2.5 nm thick, we expect the crosslinking time to be much lower than the one found for the 100-200 nm films, thus our film is in solid state after few hours of spreading at the air-D$_2$O interface.

## Adsorption at the air-water interface: effect of protein type and salt

In Figure 2, the effects of the protein type and of salt are shown for adsorption at the air-water interface. The reflectivity spectra for lysozyme and BSA at bulk concentration, C$_p$ = 1 mg·mL$^{-1}$ in pure D$_2$O and in D$_2$O at an ionic strength of 155 mM NaCl are shown in Figure 2a. These spectra are plotted as a function of $R * q^4$ versus $q$ to emphasize the high-$q$ region and the spectra are vertically-separated for better visualization. In all cases, the curves are adequately fitted with a 1-layer model. The solid lines are best-fit curves with corresponding protein concentration profiles shown in Figure 2b. In pure D$_2$O, both proteins show very low adsorption, with volume fraction $\phi_p \approx 0.1$ and thickness $d \approx 20$ Å. In the presence of 155 mM NaCl, adsorption is increased to $\phi_p \approx 0.16$ with an accompanying small increase in thickness for BSA to $d = 24$ Å. Similarly, for lysozyme, in presence of 155 mM NaCl, adsorption is increased to $\phi_p \approx 0.16$ while its thickness remains the same, $d \approx 20$ Å. The interfacial roughness for both proteins is $\sigma = 3$ Å.

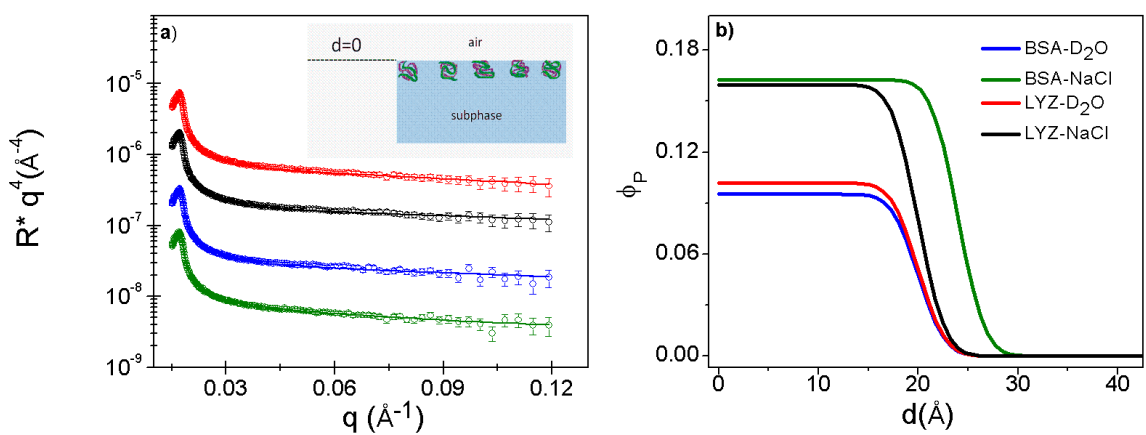

**Figure 2:** Adsorption of proteins at the air-water interface: effect of protein type and salt. (a) Neutron reflectivity plotted as a function of $R * q^4$ versus $q$: BSA in D$_2$O in the absence (blue) and presence of 155 mM NaCl (green), and lysozyme in D$_2$O in the absence (red) and presence of 155 mM NaCl (black); the curves are separated vertically for clearer viewing. The solid lines are best-fit curves with corresponding volume fraction profiles shown in (b). The protein bulk concentration, $C_p$ = 1 mg·mL⁻¹.

## Adsorption at the oil-water interface: effect of protein type and salt

For these experiments, the oil was spread on the surface of the protein solution immediately after the protein solution was introduced into the sample cell. The reflectivity curves for the adsorption of BSA and lysozyme in the presence of the nanometer-oil layer are shown in Figure 3a. The solid lines are best-fit calculated curves taking into consideration the spread oil layer. Interestingly, the fitted scattering length densities for this oil layer do not vary significantly, with an average fitted value of 0.61 ×10⁻⁶ Å⁻² which is very close the calculated value for pure ICO (0.57 ×10⁻⁶ Å⁻²). The thickness of the layer is found to vary from about 21 to 25 Å, a variation that may be due to spreading error.

The corresponding protein volume fraction profiles extracted from the fitted layers below the oil layer are given in Figure 3b. In these cases, the concentration profile starts at the oil-solvent interface ($d$ = 0), as depicted in the schematic in Figure 1c. For both proteins, a 1-layer model is inadequate to describe the adsorbed layer: in D₂O, a 2-layer model is necessary and in the presence of 155 mM NaCl a 3-layer model is required to give adequate fit. Here, $\sigma$ = 3 - 5 Å for the second layer, and 5 - 10 Å for the third diffuse layer. From these concentration profiles, the marked effect of salt can be seen clearly from the increase in the volume fraction as well as the total thickness of the adsorbed layer.

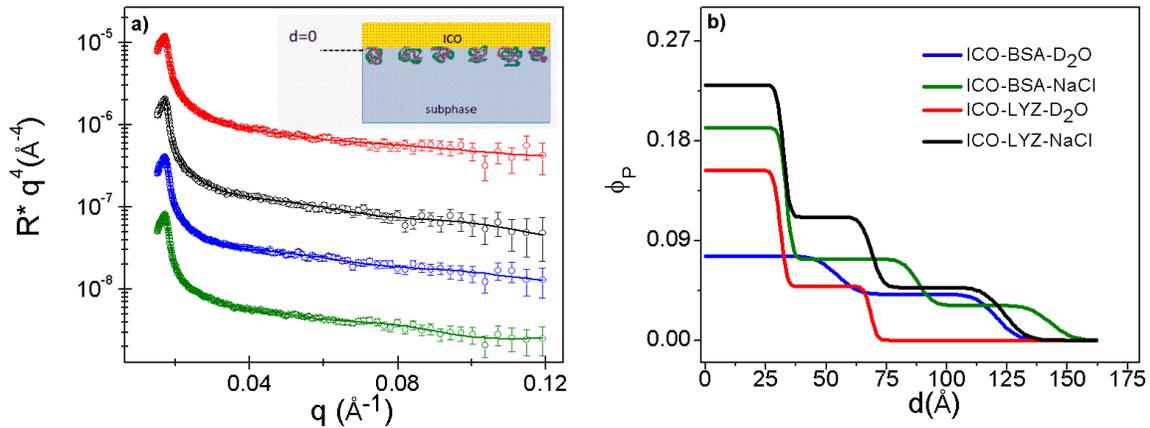

**Figure 3:** Adsorption of proteins at the (ICO) oil-water interface: effect of protein type and salt. (a) Neutron reflectivity plotted as a function of $R * q^4$ versus $q$: BSA in D₂O in the absence (blue) and presence of 155 mM NaCl (green), and lysozyme in $D_2O$ in the absence (red) and presence of 155 mM NaCl (black); the curves are separated vertically for clearer viewing. The solid lines are best-fit curves with corresponding volume fraction profiles shown in (b). The protein bulk concentration, $C_p = 1$ mg·mL$^{-1}$.

## Adsorption at the oil-water interface: effect of protein concentration

The effect of protein concentration at the oil-water interface in the presence of salt was studied by increasing the bulk protein concentration from $C_p = 1$ mg·mL$^{-1}$ to 5 mg·mL$^{-1}$. Figure 4a shows that for BSA, the reflectivity curves for the two concentrations almost superpose, with only some minor difference at high-$q$ related to structural reorganization within the layer, as shown in the concentration profile in Figure 4b. The total adsorption density remains unchanged at about $1.7 \pm 0.2$ mg·m$^{-2}$. For lysozyme (Figure 4c), the difference between the two concentrations appears slightly larger, with the corresponding concentration profiles showing some reorganization within the layer (Figure 4d). Nevertheless, the total adsorption densities remain comparable: $2.0 \pm 0.2$ mg·m$^{-2}$ at $C_p = 1$ mg·mL$^{-1}$ and $1.8 \pm 0.2$ mg·m$^{-2}$ at $C_p = 5$ mg·mL$^{-1}$.

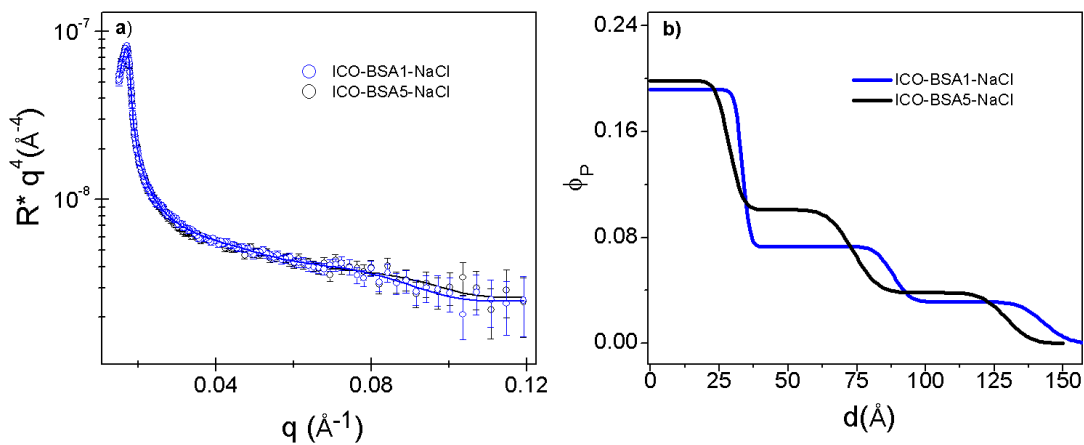

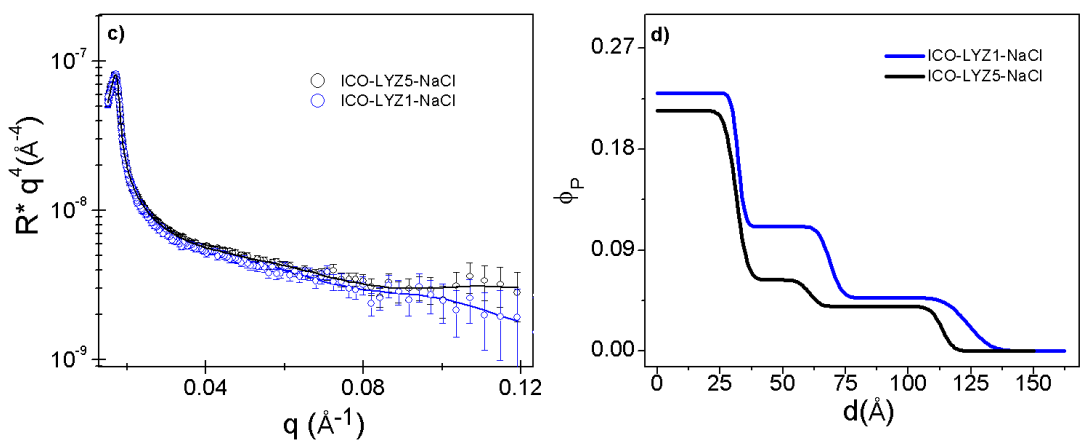

**Figure 4 :** Adsorption of proteins at the (ICO) oil-water interface in the presence of 155 mM NaCl: effect of bulk protein concentration. Neutron reflectivity plotted as a function of $R * q^4$ versus $q$ for BSA (a) and lysozyme (c) at $C_p = 1$ mg·mL$^{-1}$ (blue) and 5 mg·mL$^{-1}$ (black); the solid lines are best-fit curves with corresponding volume fraction profiles shown in (b) and (d) respectively.

In Figure 5, the concentration profiles for each protein are consolidated to facilitate a comparison of adsorption at the air-D$_2$O and oil-D$_2$O interface. The difference in adsorption behavior at the different interfaces is significant: at the air-D$_2$O interface, adsorption is low, and the concentration profile can be described by a 1-layer model, while at the oil-D$_2$O interface a 2 to 3-layer model is required to adequately characterize the protein layer that extends to a total thickness that indicates multilayer adsorption. Therefore, at the oil-D$_2$O interface, both the packing density and the total adsorbed layer thickness are increased. The enhancement by salt is also clear for both proteins and at both interfaces: the presence of 155 mM NaCl increases the volume fraction in the first layer by about 1.5 - 2 times. Interestingly, for BSA at the oil-D$_2$O interface in the absence of salt, a slight dilution/expansion (compared to the air-D$_2$O interface) of the first layer is observed. This effect may be due to repulsion between the negatively-charged protein and the residual negative charges on the silylated oil. The origin of these negative charges demonstrated by zetametry on emulsified ICO (typical value -0.98 mV) is due to the presence of free ricinoleic acids (pKa 4.99) under carboxylate form at pH 7. This oil-protein electrostatic interaction is small compared to protein-protein repulsion, since the addition of 155 mM NaCl increases adsorption at this oil interface for both the negatively-charged BSA and the positively-charged lysozyme – the adsorption density increases by about two-fold in both cases. Regarding the bulk concentration effect, increase in C$_p$ from 1 to 5 mg·mL$^{-1}$ shows some reorganization in the adsorbed layer without significant difference in the total adsorbed amount: for BSA, $\Gamma = 1.7 \pm 0.2$ mg·m$^{-2}$ at both protein concentrations; for lysozyme, the adsorption density is comparable, $\Gamma = 2.0 \pm 0.2$ mg·m$^{-2}$ and $1.8 \pm 0.2$ mg·m$^{-2}$ at $C_p = 1$ mg·mL$^{-1}$ and 5 mg·mL$^{-1}$ respectively. For both proteins therefore, a five-fold increase in bulk concentration studied here shows no further increase in adsorption density, suggesting that the maximum protein adsorption has been identified, at physiological ionic strength. All the fitted structural parameters are given in Table 1.

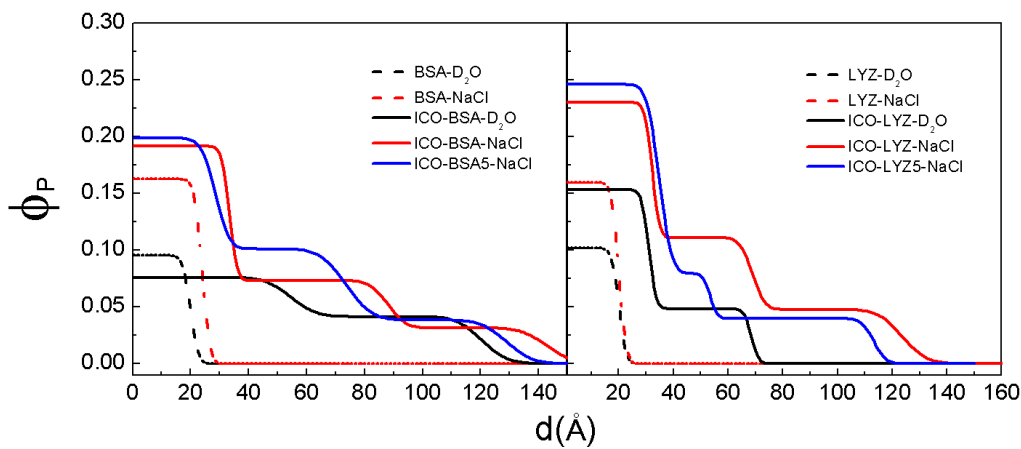

**Figure 5:** Adsorption density profiles at the air-water (dashed lines) and (ICO) oil-water (solid lines) interface, for BSA (left panel) and lysozyme (right panel): in D$_2$O at $C_p = 1$ mg·mL$^{-1}$ (black); in D$_2$O (155 mM NaCl) at $C_p = 1$ mg·mL$^{-1}$ (red) and at $C_p = 5$ mg·mL$^{-1}$ (blue).

**Table 1:** Fitted structural parameters for protein adsorbed layers at the air-water and ICO-water interface measured in D$_2$O in the absence and presence of 155 mM NaCl, at protein concentrations, $C_p = 1$ mg·mL$^{-1}$ and 5 mg·mL$^{-1}$ at pD = 7. $d$ and $\phi_p$ are the thickness and protein volume fraction in layer-$n$; $\Gamma$ is the total adsorption density.

| $C_p$ | Interface | layer $n$ | $d \pm 2$ (Å) | $\phi_p$ | $\Gamma$ (mg·m$^{-2}$) |
|---|---|---|---|---|---|
| | **BSA** | | | | |
| 1 mg·mL$^{-1}$ $1.5 \cdot 10^{-5}$ M | air-D$_2$O | 1 | 20 | 0.095 | $0.27 \pm 0.05$ |
| | air-D$_2$O (155 mM NaCl) | 1 | 24 | 0.163 | $0.55 \pm 0.07$ |
| | ICO-D$_2$O | 1 | 55 | 0.076 | $0.96 \pm 0.15$ |
| | | 2 | 66 | 0.041 | |
| | ICO-D$_2$O (155 mM NaCl) | 1 | 33 | 0.192 | $1.70 \pm 0.22$ |
| | | 2 | 55 | 0.073 | |
| | | 3 | 55 | 0.032 | |
| 5 mg·mL⁻¹ 7.6·10⁻⁵ M | ICO-D₂O (155 mM NaCl) | 1 | 29 | 0.199 | 1.73 ± 0.22 |
| | | 2 | 45 | 0.101 | |
| | | 3 | 56 | 0.037 | |
| **Lysozyme** | | | | | |
| 1 mg·mL⁻¹ 6.9·10⁻⁵ M | air-D₂O | 1 | 20 | 0.102 | 0.29 ± 0.05 |
| | air-D₂O (155 mM NaCl) | 1 | 20 | 0.160 | 0.46 ± 0.07 |
| | ICO-D₂O | 1 | 31 | 0.153 | 0.94 ± 0.13 |
| | | 2 | 37 | 0.049 | |
| | ICO-D₂O (155 mM NaCl) | 1 | 32 | 0.230 | 2.02 ± 0.24 |
| | | 2 | 37 | 0.111 | |
| | | 3 | 55 | 0.048 | |
| 5 mg·mL⁻¹ 3.5·10⁻⁴ M | ICO-D₂O (155 mM NaCl) | 1 | 35 | 0.246 | 1.79 ± 0.23 |
| | | 2 | 19 | 0.079 | |
| | | 3 | 60 | 0.040 | |

At equilibrium, during protein adsorption at fluid interfaces the following steps take place: firstly, the molecules diffuse from the bulk phase to the interface, and this is accompanied by an increase in the surface pressure. Then conformational changes may take place, sometimes also denaturation, and at the final stage multilayers may form. Globular proteins contain hydrophilic charged amino acid groups which are distributed in the outer shell of the protein and during adsorption at the air-water interface the hydrophobic groups are expected to be exposed to the air while the hydrophilic ones towards the water. The packing of the proteins is known to be driven by electrostatic, hydrophobic, van der Waals, entropic and steric interactions.[21, 30-33] However, our knowledge of protein conformations at fluid interfaces is still limited and the denaturation of proteins has not been understood.

In this work, the adsorption densities of two proteins, BSA and lysozyme, under otherwise identical experimental conditions, i.e. in the absence and presence of salt, and in the absence and presence of ICO have been compared. At the air-water interface, adsorption of both charged proteins is low: addition of 155 mM NaCl produces a 1.5 to 2-fold increase in adsorption but nevertheless yields $\Gamma < 1$ mg·m⁻². Compared to the literature, these results are lower than as those reported, for example by Lu et al. under similar pH and temperature conditions.[14, 27-28] The reason for this contrast is not clear. We point out however that for the protein solutions used in our current study, apart from the NaCl added to control the ionic strength, no other chemical or buffer was added. Furthermore, the solutions were freshly-prepared, refrigerated, and used within two days. It is not clear how or if any at all, buffer chemicals such as phosphates used in the reported studies or other commonly-used buffers such as Tris (2-Amino-2-(hydroxymethyl)propane-1,3-diol), affect the protein structure, although binding of Tris to proteins and changes in their activity have been reported.[34-35] The influence of buffer nature as well as its concentration was highlighted by Zhang et al. who showed that lysozyme which is a positively charged macromolecular system at neutral pH, precipitates in agreement with an inverse Hofmeister behavior only at low salt concentrations, but reverts to a direct Hofmeister series when the salt concentration increases.[36] Medda et al. studied the specific-ion binding of BSA at a physiological concentration of 0.1 M using potentiometric titration and electrophoretic light scattering and they reported that anions bind to the protein surface at an acidic pH, where the protein carries a positive charge, and at the isoelectric point, in accordance with the Hofmeister series ($Cl^- < Br^- < NO_3^- < I^- < SCN^-$).[37] Regarding the age of the solution, we have found that BSA solution aged one week or more gave a higher adsorption density, suggesting that the protein may have undergone some denaturation. Comparison of results from different studies are therefore not always direct and absolute. We hereby compare and contrast the adsorption behavior of BSA and lysozyme at the air-water and the oil-water interface from the same freshly-prepared solutions.

At the oil-water interface in the absence of salt, the adsorption density is increased to about 1 mg·m$^{-2}$ for both proteins. Note that while the total adsorption density is similar, a difference can be seen in the packing density in the first layer of the profile (oil-side): for BSA, it is more dilute and expanded (compared to the air-water interface), but for lysozyme it is more concentrated and the profile more well-defined. This may be an effect of the residual negative charges on the silylated oil causing a small degree of repulsion between the oil and BSA. The presence of 155 mM NaCl increases the adsorption density from 1 mg·m$^{-2}$ to 1.7 mg·m$^{-2}$ for BSA and to 2.0 mg·m$^{-2}$ for lysozyme. The oil-protein electrostatic interaction is therefore small compared to protein-protein repulsion, since the addition of NaCl increases adsorption of both the negatively-charged BSA and the positively-charged lysozyme; furthermore, the degree of impact of the NaCl is similar at both the air-water and the oil-water interface, giving rise to a two-fold increase compared to the absence of salt. The screening of electrostatic repulsion between proteins by the addition of salt leading to increase in adsorption density is clear from our reflectivity measurements. Notably, the concentration profiles of the adsorbed layers reveal increased packing density as well as multilayer formation at the oil-water interface. Comparing our results with casein, a random coil protein, Dickinson et al. found that the adsorption of casein at both air/$D_2O$ and hexane/$D_2O$ interfaces can be described by a dense inner layer of 2 nm thickness and a secondary layer of thickness 5-7 nm.[38] The latter extends into the aqueous phase, with a volume fraction in the range of 0.15 to 0.2. Interestingly, the authors did not report any significant increase of the adsorption in presence of oil, like in the present study. Other studies have also shown the particular impact of protein charge on adsorption at neutral oil-water interfaces. These include BSA reaching optimum adsorption density of 3.2 mg·m$^{-2}$ under conditions of reduced protein-protein repulsion at the isoelectric point at the hexadecane-water interface,[18] and soy protein adsorption at soybean oil-water interface at elevated ionic strength.[39] The difference in adsorption behavior at the air-water and oil-water interface is significant and the impact of oil is marked for all cases investigated. Enhanced protein adsorption in the presence of oil can be attributed to attractive hydrophobic and van der Waals, in particular dispersion, forces. According to Sengupta et al.[40] dispersion interactions play a primary role in dictating protein adsorption at the different interfaces. They find from potential energy calculations that dispersion interaction between proteins and oil is always attractive, while the dispersion interaction between proteins and water is always repulsive. Thus, although the sum total of all the interaction energies is attractive at both air-water and oil-water interfaces, the dominating repulsive dispersion part renders the van der Waals forces less attractive at the air-water interface compared to the oil-water interface.

A comparison of the concentration profile and the thickness of the adsorbed layer with the dimension of the protein may indicate the orientation of the adsorbed molecule. Both the globular proteins in aqueous solutions are believed to have an ellipsoidal structure. For BSA, the most commonly-cited dimension is $40 \times 40 \times 140$ Å$^3$ as measured by small-angle neutron scattering by Chen et al.[41], although more recently a triangular oblate structure has been proposed, with dimension $32 \times 84 \times 84$ Å$^3$ from hydrodynamic modeling by Ferrer et al.[42] and dimension $26 \times 80 \times 80$ Å$^3$ from neutron scattering by Kundu et al.[43] For lysozyme, the commonly-accepted structure is the crystalline dimension $30 \times 30 \times 45$ Å$^3$,[44] comparable with the dimension in solution of $28 \times 28 \times 50$ Å$^3$ measured by X-ray scattering.[45] At the air-water interface, the adsorption density obtained in the current study is low and the simplest 1-layer model yields an adsorbed layer of thickness 20 - 24 Å for BSA (in absence and presence of NaCl respectively) and 20 Å for lysozyme, suggesting a sideways orientation with the long axis parallel to the surface. For BSA, the layer thickness is thus closer to the dimension of the minor-axis of the oblate structure propose by Kundu et al. and by Ferrer et al. Nevertheless, the thicknesses of the adsorbed layer for both proteins are smaller than their sideways dimensions. One possible reason for this result is the protrusion of the adsorbed molecule into the air-phase. Using different contrast-variations in their neutron reflectivity measurements to evaluate separately the portion of the adsorbed molecule in the air and in the water phase, Lu et al.[14, 27] found that the part of the BSA layer that protrudes into the air can vary from 5 - 10 Å, with no particular trend in the variation with respect to solution conditions; for lysozyme, this protruded layer can be as high as 15 Å due to its higher molecular rigidity.[28] In our studies, due to the low adsorbed amount, the part that protrudes into the air does not appear to contribute sufficiently to the total signal, therefore the thickness that is reported here reflects mostly the submerged part of the adsorbed layer. A second possible explanation is a structural deformation of the molecule upon adsorption - a flattening of the adsorbed molecule in particular under low adsorption and low lateral repulsion conditions.[14] At the oil-water interface, the increased adsorbed amount is modeled with a multilayer-profile. Here, in contrast to the air-water interface, the thickness of the first layer (oil-side) compares well with the dimension of the protein: 29 - 33 Å for BSA (in the presence of NaCl where charged surface-protein repulsions may be neglected, and with reference to the oblate dimension proposed by Kundu et al. and Ferrer et al.), and 31 - 35 Å for lysozyme. These results suggest a sideways orientation in the first adsorbed layer; in this case, the fact that the thickness of the primary adsorbed layer compares well with the protein dimension suggests complete immersion of the molecule in the water phase. The oil phase thus appears less penetrable to the protein molecule than the air-phase, an observation also supported by the scattering length density of the oil layer that remains mostly unchanged upon protein adsorption. Zare et al. showed that the penetration of the proteins in the oil phase depends on the hydrophobicity of the oil. In line with our results, β-lactoglobumin adsorbed on a very hydrophobic oil, decane, did not penetrate the oil, staying flattened onto the oil surface.[46] In our case, the ensuing layers below the primary layer are larger than the dimensions of the minor axes of the proteins. In particular, the last layer at the subphase-side is dilute and diffuse, extending to as high as an average of 60 Å. Although these thicknesses approach the dimensions of the long axes of the proteins, note however, as mentioned before, that the fitted "*n*-layer" concentration profile corresponds to the scattering length density (compositional) distribution and, due to the very dilute nature of these inner layers, there is insufficient resolution to make a direct correspondence to the exact molecular positioning within the total adsorbed layer. Therefore, it is uncertain whether the increased layer thickness reflects a change in molecular orientation from sideways to longways (long axis perpendicular to the surface), or a stacked sideways orientation. We focus our analysis therefore on the total adsorbed amount of the protein layers. Figure 6 shows a schematic of their approximate adsorbed orientation.

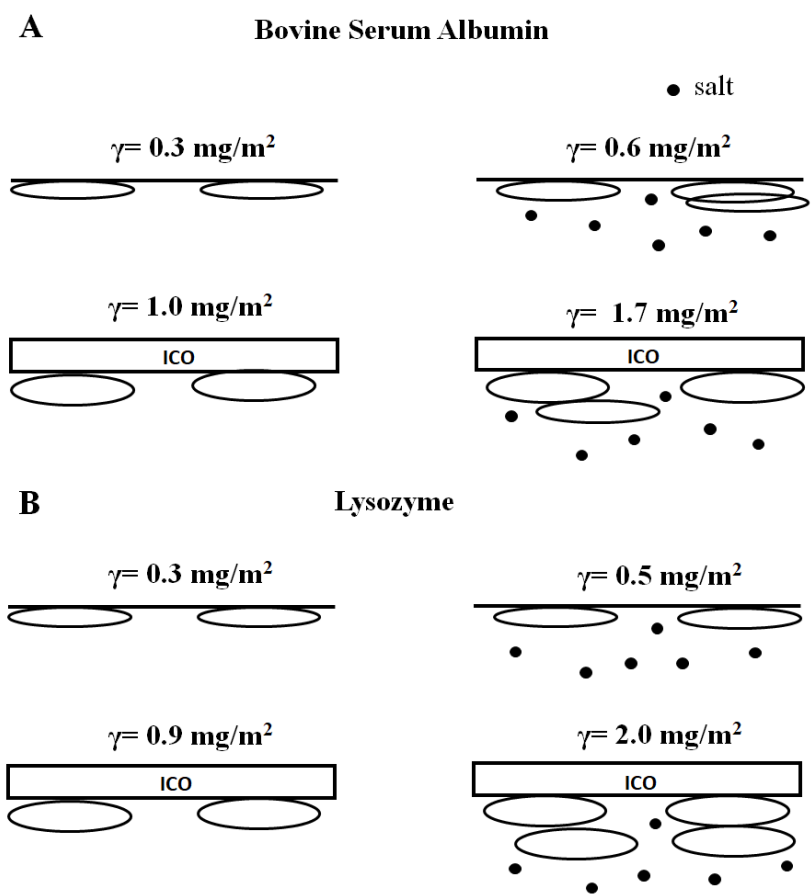

**Figure 6:** Schematic summary of protein adsorption of BSA (a) and lysozyme (b) at the air-water (top) and ICO-water (bottom) interface, in the absence (left) and presence of salt (right, represented by dots). The figure is intended to show only an approximate representation of the orientation of the adsorbed molecule; each layer in the *n*-layer model corresponds to the compositional distribution, it does not necessarily correspond to the exact molecular positioning within the total adsorbed layer.

Regarding the effect of protein bulk concentration on the adsorption density, no significant difference is obtained from a five-fold increase in bulk concentration (from 1 to 5 mg·mL⁻¹) for both proteins: $\Gamma = 1.7 \pm 0.2$ mg·m⁻² at both concentrations for BSA, and $\Gamma = 2.0$ and $1.8 \pm 0.2$ mg·m⁻² at $C_p = 1$ and 5 mg·mL⁻¹ respectively for lysozyme. The fact that the adsorption density does not increase at higher concentrations is encouraging, given that in several biomedical applications the protein concentration varies, i.e. serum albumin which is a major blood protein has a concentration of 35 - 50 mg·mL⁻¹ while the concentration of lysozyme in tears is $0.75 - 3.3$ mg·mL⁻¹ and $5\times10^{-7}$ - $4\times10^{-6}$ mg·mL⁻¹ in saliva.[47-48] In this study, the concentration of the bulk protein solution has been expressed in mg·mL⁻¹, but using molar concentrations, the result is the same, the adsorption remains virtually independent of the molar concentration (see SI, Figure S4). We have investigated protein adsorption in quasi-physiological conditions, i.e. ionic strength of 155 mM which is very close to the physiological ionic strength and at pH 7. For the purpose of this study, which is to focus on the adsorption of proteins at the oil-water interface, we have conducted the experiments at room temperature, T = 22 - 23 °C instead of the physiological temperature of 37°C. This was to minimize evaporation during the spreading of the oil on the protein solution, and to keep the same oil-spreading protocol that was used to produce thick cross-linked oil films in our past studies that were performed at room temperature. However, a small increase in adsorption is expected at higher temperatures due to the entropic nature of hydrophobic interactions.

Lastly, we compare the results of protein adsorption obtained in this study on nanometric oil films with those from our recent studies on thick cross-linked films. Large-size cross-linked ICO films of thickness ~ 200 μm can be formed at the air-water interface[7] or by spin-coating.[10] In the SI (Fig. S2) the chemical structure of the castor oil and its functionalized form ICO are shown, and the macroscopic transparent solid film obtained after the cross-link reaction of 1g of ICO deposited on a surface area of 50.3 cm² of water. Spatial characterization by wide angle X-Ray scattering (WAXS) of the film, together with the virgin castor oil and the functionalized oil ICO, shows an amorphous structure with a strong peak at $q = 1.4$ Å⁻¹ that decreases in intensity with functionalization and cross-linking, with no significant shift in the peak position. This scattering feature is related to the presence of a short-range local order of the triglycerides in coexistence with a disordered amorphous phase of the polyurethanes. The Young's modulus and hardness of the film, in the MPa range, are tunable by the ratio of the silica precursor (coupling agent) to castor oil. In all cases, the contact angle of the solid film is below 90°. Adsorption studies of BSA and lysozyme on spin-coated cross-linked ICO films of thickness of about 200 nm were carried out using the quartz crystal microbalance with dissipation (QCM-D) method and atomic force microscopy (AFM). In that study, protein solutions were flowed over the ICO film at a rate of 20 μL·min$^{-1}$ for four hours. After these experiments, the films were stored in physiologic buffer salt solutions and AFM was performed to measure the thickness of the adsorbed protein. The QCM-D measurements indicate that BSA and lysozyme, in physiological buffered salt solutions, form rigid layers on the film surface with adsorbed densities close to 3 mg·m$^{-2}$. The present neutron reflectivity results, $1.7 \pm 0.2$ mg·m$^{-2}$ for BSA and $2 \pm 0.2$ mg·m$^{-2}$ for lysozyme, are therefore comparable to the QCM-D measurements. We exclude that the difference in the exact adsorbed amount is due to the nature of the spin-coated films, because we saw with the IR experiments that the degree of cross-linking in the silylated oil after 8 hours of its deposition is very high. In fact, the ICO film is in the solid state and is found to remain stable and devoid of significant protein penetration. The QCM-D and AFM measurements yielded adsorbed layer thicknesses in the range 65 - 83 Å for BSA and around 80 Å for lysozyme. Inspection of the concentration profiles from the neutron reflectivity measurements shows that in the presence of 155 mM NaCl, the major contribution to the adsorbed layer that constitutes over 80% of the total adsorbed amount (layer-1 and layer-2 of the concentration profile) extends to a similar range in thickness, 70 - 80 Å. The good agreement of the results obtained by these three different techniques (QCM-D, AFM, Neutron Reflectometry) leads us to identify and ascertain that the optimal protein adsorption at the ICO-water interface falls in the range of 2 - 3 mg·m$^{-2}$.

**CONCLUSIONS**

Protein adsorption at the solid-liquid interface is closely related to the biocompatibility of a new material. The long-term use of contact lenses and devices that are in contact with blood is tested according to the adsorbed amount of proteins. Thus, understanding the protein adsorption behavior is crucial for the evaluation of a new bio-material. In this study, castor oil molecules have been functionalized (ICO) with a silylated cross-link agent to form a vegetable oil-based hybrid material which can be used in biomedicine as artificial biomaterial, for implants and transdermal drug delivery systems. Neutron reflectivity has been employed to examine the adsorption of two model proteins, bovine serum albumin and lysozyme at the ICO-water interface. The ICO in this case is a nanofilm spread on the water or on the protein solution surface. The effects of salt with an ionic strength close to the physiological concentration and of protein bulk concentration have been investigated. The key result is the identification of the most favorable and unfavorable conditions for adsorption of BSA and lysozyme on the ICO nanofilms. The adsorption densities, under conditions of screened protein-protein repulsion by salt at ionic strength close to physiological conditions, are found to remain almost constant with the protein concentrations studied, allowing us to identify the optimum protein adsorption on these films. Notably, adsorption is found to be non-specific, and at the ICO-water interface is three to four times higher than at the air-water interface. We attribute this increase in affinity to the attractive hydrophobic and van der Waals forces, in particular the dispersion forces, shown by Sengupta et al. [40] to be always positive between proteins and oil and repulsive between proteins and water. Thus, although the sum of all the interaction energies is attractive at both air-water and oil-water interfaces, dispersion interactions play a primary role in governing protein adsorption at the different interfaces. The volume fraction profiles suggest that in the principal layer (oil-side), the proteins adsorb with their long axis parallel to the surface. The presence of salt in the subphase increases the packing density in this layer and engenders multilayer adsorption. We have seen that protein adsorption is controlled by the water activity and its structure at the surface.[49] In our study, we argue that the small surface coverage could be related to the low hydrophobicity of the ICO driving preferentially the system to maximize protein-protein contacts rather than protein-ICO ones, thus preventing proteins from covering the whole surface. This hypothesis would explain our results and would lead to multilayer formation. This is also corroborated by the enhanced adsorption observed after addition of salt due to screening of electrostatic interactions promoting protein aggregation, and thus multilayering. Importantly, the results obtained on these nanofilms are comparable to the results obtained on thick cross-linked ICO films investigated by quartz crystal microbalance with dissipation and atomic force microscopy. The ensemble of the results obtained by these different techniques (Neutron Reflectometry, QCM-D, AFM), on very thin nanometer-films and on intermediate thickness hundred-nanometer films, leads us to identify the optimal conditions and quantify the protein adsorption density at the ICO-water interface to fall in the range of 2 - 3 $mg \cdot m^{-2}$. It is hoped that this study provides a starting point for future investigations of eco-friendly vegetable oil hybrid materials of relevance in several recent biomedical applications, replacing conventional petroleum-based polymeric films in the pharmaceutical industry.

**Acknowledgements:** We thank the French Neutron Federation F2N and the French national neutron scattering laboratory, Laboratoire Léon Brillouin (CEA - Saclay) for beam time. The authors gratefully acknowledge the financial support of this work by Labex CheMISyst and University of Montpellier (UM).

**Supporting Information:** A sketch showing proteins adsorbed at the interfaces and the different parameters studied in this work; previously published data of thick silylated castor oil films; ATR spectra of silylated castor oil film at different times; protein adsorption as a function of molar bulk concentration.

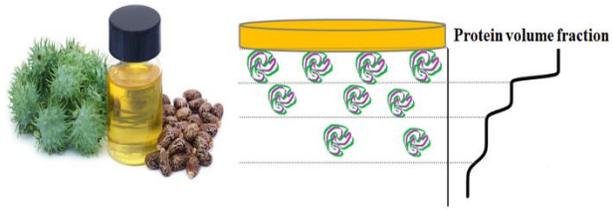

TOC: Protein adsorption of proteins at the silylated castor oil-water interface